%% file: u1paper_v3.tex
\documentclass[12pt,a4paper]{article}
\usepackage{amssymb}
\usepackage{amsmath,amsthm}
\usepackage{amsfonts}
\usepackage{epsfig}
\usepackage{color}
\usepackage{hyperref}
\usepackage{bm}
\outer\def\beginsection#1\par{\medbreak\bigskip
      \message{#1}\leftline{\bf#1}\nobreak\medskip
\vskip-\parskip
      \noindent}
\newcommand{\eq}{\begin{equation}}
\newcommand{\eqx}{\end{equation}}
\newcommand{\eqn}{\begin{eqnarray}}

\newcommand{\e}{{\rm e\,}}

\newcommand{\R}{\mathbb{R}}

\newcommand{\cT}{{\cal T}}

\def\be#1\ee{\begin{equation}#1\end{equation}}
\def\bea#1\eea{\begin{align}#1\end{align}}

\newcommand{\bi}{\begin{itemize}}

\newcommand{\eqnx}{\end{eqnarray}}
\newcommand{\ei}{\end{itemize}}

\newcommand{\nn}{\nonumber}
\newcommand{\ra}{\rangle}
\newcommand{\la}{\langle}

\title{
Positive Representations of a Class of Complex Measures
}

\author{
Erhard Seiler$^a$\footnote{email: ehs@mpp.mpg.de} 
and 
Jacek Wosiek$^b$\footnote{email: Jacek.Wosiek@uj.edu.pl} \;\; 
\mbox{} \\
\mbox{} \\
 $^a${\em\normalsize Max-Planck-Institut f\"ur Physik 
(Werner-Heisenberg-Institut)}\\
{\em\normalsize M{\"u}nchen, Germany} \\
 $^b${\em\normalsize M. Smoluchowski Institute of Physics, 
Jagiellonian University}\\
{\em\normalsize Cracow, Poland}
}
\begin{document} 

\maketitle

\begin{abstract} \normalsize\noindent 
We study the problem of constructing positive representations of complex 
measures. In this paper we consider complex densities on a direct product 
of $U(1)$ groups and look for representations by probability distributions 
on the complexification of those groups. After identifying general 
necessary and sufficient conditions we propose several concrete 
realizations. Finally we study some of those realizations in examples 
representing problems in abelian lattice gauge theories. \end{abstract}

\hskip4mm Keywords: Lattice Field Theory, Complex Actions, Wilson Loops, 
Polyakov Loops

\newpage
\input sec1

\input sec2

\input sec3

\input sec4

\input references
\end{document}

%% file: sec1.tex
\section{Introduction}

Feynman's ``sum over histories'' assigns complex amplitudes rather than
positive probabilities to the different possible histories, so expectation 
values are obtained by complex weights. Examples range from simple time 
dependent problems in quantum mechanics to quantum field theory. 
This feature  persists in many 
cases even when continuing to imaginary (euclidean) time.

On the other hand, existing positive representations of euclidean Feynman 
path integrals provide an intriguing statistical interpretation of many 
field theories and have led to the development of very successful, 
non-perturbative, lattice solutions of QCD and other theories 
\cite{wilson, creutz, weisz, liu}.

The most prominent situations where such representations are still 
lacking, however, include motion in an external magnetic field, detailed 
studies of the mechanism of confinement \cite{Fuk, HW, ichie}, QCD at 
finite chemical potential \cite{FiniteMu} and the original formulation of 
fermionic path integrals in terms of Grassmann variables \cite{Be,hybrid}. 
Similarly, studies of the real (i.e. Minkowski) time evolution of quantum 
systems encounter the same difficulty \cite{berges1,berges2}. Technically 
the problem, often referred to as the ``sign problem'', arises because 
summation of the oscillating functions of the huge number of variables 
cannot be done using statistical, i.e. Monte Carlo techniques.

Consequently, research on the sign problem has a long history which goes 
back to the beginning of the formulation of lattice field theory. One way 
to deal with the difficulty employs stochastic quantization based on the 
the complex Langevin equations \cite{P,Kl}. A revival of the interest in 
the sign problem was triggered by recent progress \cite{berges1, berges2} 
and \cite{aarts_stam}, leading to successes in more and more realistic 
models up to full QCD \cite{S1,S2, sexty, expansion}.

Nevertheless the old troubles \cite{AY,HW2}, which had plagued the method, 
resurfaced again and, in spite of substantially better understanding 
\cite{S4}, the approach still has serious difficulties and limitations 
\cite{Bl,Ph,aarts2017}.

The essential goal of the complex Langevin approach is summarized by the 
following relation 
\eqn 
\frac{\int f(x) \rho(x) dx}{\int \rho(x) dx } = \frac{\int 
\int f(x+i y) P(x,y) dx dy }{\int \int P(x,y) dx dy}\,, 
\label{me} 
\eqnx 
where the weight $\rho(x)=e^{-S(x)}$ can be complex, and $P(x,y)$ is the 
equilibrium probability distribution the stochastic process in the 
complexified configuration space, associated with the analytically 
continued action $S(x+iy)$. The precise form of the Langevin equation is 
not relevant here. It suffices to say that for a real action the Langevin 
process is real -- the density $P(x)$ is concentrated on the real axis. It 
satisfies the Fokker-Planck (FP) equation whose solution converges to 
$e^{-S(x)}$ for large Langevin time. On the other hand, for complex 
actions the stochastic trajectory is driven into the complex extension of 
the configuration space. The density $P(x,y)$ satisfies a FP equation in 
twice as many variables, however the asymptotic (in the Langevin time) 
behaviour of solutions and their equilibrium distribution $P$ are in 
general not known explicitly and their relation to the original, complex 
actions is  quite intricate \cite{DH}.

One can also attempt to construct a positive density $P(x,y)$ directly 
from the ``matching equation'' eq.(\ref{me}). It was proven in \cite{Wei} 
that such distributions indeed exist under certain conditions, however no 
practical construction was provided. Explicit derivations for Gaussian and 
related cases were given  in Ref.\cite{Sa1}; they were extended 
to densities on certain compact spaces in \cite{Sa2}. Recently Salcedo 
\cite{Sa3,Sa4} presented very nice explicit derivations for a circle. The  
general constructions discussed here will contain his solution as a 
special case. A somewhat different approach was proposed recently, 
avoiding again any reference to stochastic processes \cite{Wo}. With the 
aid of a second complex variable a simple integral relation between $\rho$ 
and $P$ was derived and solved for the Gaussian case again. This time, 
however, the generalization to an infinite number of degrees of freedom, 
including careful treatment of the continuum limit, was also achieved. 
This provided for the first time a positive representation of some classic 
quantum mechanical problems directly in the Minkowski time.
 
In this paper we construct positive representations for measures on a 
class of compact abelian group spaces. Our approach was already extended to non-compact variables by Ruba and Wyrzykowski \cite{RW}.  For an interesting, entirely new attempt see Ref.\cite{WR}. 

Evidently the solution of 
eq.(\ref{me}) in terms of $P$ is not unique and we present a few 
realizations which illustrate this freedom. The general construction, 
given in the next Section, is then followed by few examples with 
various numbers of variables with the aim of finding a positive 
representation of Wilson lines in $U(1)$ lattice gauge theory.

%% file: sec2.tex
\section{General construction for $U(1)$ models}

\subsection{The principle}

We describe the general priniciple by which positive measures, equivalent 
to a complex measure on the $N$-torus $\cT^N=(U(1))^N$ can be constructed. 
The starting point is a complex density $\rho$, representing a complex 
measure on $\cT^N=(U(1))^N$, normalized as  
\be
\int_{\cT^N} \rho(\vec x)d^Nx =1\, 
\label{norm}
\ee
and of bounded variation, i.e.
\be
\int_{\cT^N} |\rho(\vec x)|d^Nx\equiv B < \infty\,. 
\label{boundvar}
\ee
Unlike in the Complex Langevin approach, we do not need to assume that it 
has any holomorphicity properties; in fact $\rho$ may contain $\delta$ 
functions.

The weight $\rho$ has the Fourier decomposition 
\be
\rho(\vec x)=\sum_{\vec n} a_{\vec n} \exp\left(i\vec n\cdot \vec x 
\right);\quad  a_{\vec n} =\frac{1}{(2\pi)^N}\int_{\cT^N} \rho(\vec x) 
\exp\left(-i\vec n \cdot \vec x\right) d^Nx
\ee
and by eq.(\ref{boundvar}) the Fourier coefficients are uniformly 
bounded 
\be
|a_{\vec n}|\le 2\pi B\,.
\label{cont}
\ee
Note that by eq.(\ref{norm}) $a_{\vec 0}=1$. 
We are going to construct a probability measure $P(\vec x, \vec y)$ on the 
complexification $\cT_C^N=\cT^N\times \R^N$ of $\cT^N$ in terms of its partial 
Fourier modes 
$P_{\vec n}(\vec y)$:
\be
P_{\vec n}(\vec y) = 
\int_{T^N}P(\vec x, \vec y) \exp\left(-i\vec n\cdot\vec x\right)d^Nx\,.
\ee
Consistency of $P$ with $\rho$ means that the expectation values agree for 
holomorphic obervables, in particular that 
\be  
\frac{1}{(2\pi)^N}\int_{\cT^N_C} \e^{-i\vec n\cdot \vec x+\vec n\cdot \vec 
y}P(x,y) d^Nx d^Ny
= \frac{1}{(2\pi)^N} 
\int_{\cT^N_C} \e^{-i\vec n\cdot \vec x} \rho(\vec x) d^Nx= a_{\vec n}
\label{consist1}
\ee
or equivalently
\be
\int_{\R^N} P_{\vec n}(\vec y) \e^{\vec n\cdot \vec y} d^Ny=a_{\vec n}\,,
\label{consist2}
\ee
and the requirement that $P$ is real means 
\be
P_{-\vec n}(\vec y)=P_{\vec n}(\vec y)^* 
\label{reality}
\ee
for all $\vec n$. 
Since the full density $P$ is given by
\be 
P(\vec x, \vec y)=\sum_{\vec n} P_{\vec n}(\vec y)\exp\left(i\vec n\cdot 
\vec x \right)\,,  
\label{fourier}
\ee
positivity of $P$ will be assured if the zero mode $P_{\vec 0}(\vec y)$ is 
positive and dominates the other modes sufficiently, so typically some 
damping of the higher modes will be needed.

It is useful to rephrase the conditions in terms of the Fourier 
coefficients of the real and imaginary parts of $\rho$. They are, 
respectively
\be
\alpha_{\vec n}\equiv \frac{1}{2}(a_{\vec n}+a_{-\vec n}^*)\,, \qquad 
\beta_{\vec n}\equiv \frac{1}{2i} (a_{\vec n}-a_{-\vec n}^*)\,, 
\ee
and satisfy
\be
\alpha_{-\vec n}=\alpha_{\vec n}^*\,, \qquad
\beta_{-\vec n}=\beta_{\vec n}^*\,.
\ee
In particular, because of the normalization of $\rho$ we have
\be
\alpha_{\vec 0}=1\,, \qquad \beta_{\vec 0}=0\,.
\ee
Taking the complex conjugate of eq.(\ref{consist2}) and replacing $\vec n$ 
by $-\vec n$, we obtain
\be
\int P_{\vec n}(\vec y) \e^{-\vec n\cdot \vec y}
d^Ny=a_{-\vec n}^*\,,
\ee
so that the consistency condition eq.(\ref{consist2}) is equivalent to the 
two conditions
\be
\int P_{\vec n}(\vec y) \cosh(\vec n \cdot \vec y)d^Ny=
\alpha_{\vec n}\,, \qquad 
\int P_{\vec n}(\vec y) \sinh(\vec n \cdot \vec y)d^Ny=
i\beta_{\vec n}\quad\forall \vec n\,,
\label{consist3}
\ee 
or, defining the even and odd parts of $P_{\vec n}$ by
\be
P^+_{\vec n}(\vec y)\equiv \frac{1}{2} \left(P_{\vec n}(\vec y)+P_{\vec 
n}(-\vec y) \right)\,;\qquad 
P^-_{\vec n}(\vec y)\equiv \frac{1}{2} \left(P_{\vec n}(\vec y)-P_{\vec    
n}(-\vec y) \right)\,,
\ee

\be
\int P^+_{\vec n}(\vec y) \cosh(\vec n \cdot \vec y)d^Ny=\alpha_{\vec n}\,, 
\qquad
\int P^-_{\vec n}(\vec y) \sinh(\vec n \cdot \vec y)d^Ny=i\beta_{\vec n}
\quad\forall \vec n\,.
\label{consist4}
\ee

These relations are the basis of our construction. Their solutions are not unique and some possibilities are discussed below in detail.

{\bf Remark:} These two conditions are necessary and sufficient for $P$ 
being a {\em real} measure equivalent to $\rho$. They also have to hold 
for the equilibrium measure of the Complex Langevin approach in order to 
give correct results.

For a real $\rho$ with a possible sign problem, $\beta_{\vec n}=0$, hence
$P$ will be even in $\vec y$; on the other hand a complex $\rho$ will 
require that $P$ has a part that is odd in $\vec y$. 

\subsection{Concrete realizations}

The consistency conditions eq.(\ref{consist3}) can be satisfied in a wide 
variety of ways. Let us consider first the following simple ansatz
\bea
P^+_{\vec n}(\vec y)=&
\frac{A_{\vec n}}{2(2\pi\sigma)^{N/2}}
\left[\exp\left(-\frac{(\vec y-\vec y_s)^2}{2\sigma}\right) +
\exp\left(-\frac{(\vec y+\vec y_s)^2}{2\sigma}\right)\right]\notag\\
P^-_{\vec n}(\vec y)=&
\frac{i B_{\vec n}}{2(2\pi\sigma)^{N/2}}
\left[\exp\left(-\frac{(\vec y-\vec y_s)^2}{2\sigma}\right) -
\exp\left(-\frac{(\vec y+\vec y_s)^2}{2\sigma}\right)\right]\,.
\label{Ansatz}
\eea
Consistency eq.(\ref{consist3}) will be fulfilled by choosing
\be
A_{\vec n}=\frac{\alpha_{\vec n}}{\cosh(\vec n \cdot \vec y_s)}
\exp\left(-\frac{\vec n^2\sigma}{2}\right)\,,\qquad
B_{\vec n}=\frac{\beta_{\vec n}}{\sinh(\vec n \cdot \vec y_s)}
\exp\left(-\frac{\vec n^2\sigma}{2}\right) 
\ee
(and of course $B_{\vec 0}=0$). Of course one has to make sure that $\vec 
n \cdot \vec y_s\neq 0$ for all $\vec n$.

To ensure that $P$ is positive, the damping of the higher modes have to be 
so strong that the zero mode dominates the sum of the others. This can be 
achieved either by making $\sigma$ or $y_s$ large enough; see, however, 
below about avoiding the possible small denominator.

{\bf Remark:} In eq.(\ref{Ansatz}) it is allowed to choose $\sigma$ and 
$\vec y_s$ different for each $\vec n$ and also different for the even and 
odd terms. 

Defining 
\be 
\lambda_{\vec n}\equiv \frac{A_{\vec n}+iB_{\vec n}}{2}\,,\qquad
\mu_{\vec n}\equiv \frac{A_{\vec n}-iB_{\vec n}}{2}\,,
\ee
we find 
\bea
\lambda_{\vec n} =&
\exp\left(-\frac{\vec n^2\sigma}{2}\right)
\frac{\e^{\vec n\cdot\vec y_s}a_{\vec n}-
\e^{-\vec n \cdot \vec y_s}a_{-\vec n}^*} {2\sinh(2\vec n\cdot \vec
y_s)}
\notag\\
\mu_{\vec n}= &\exp\left(-\frac{\vec n^2\sigma}{2}\right)
\frac{\e^{\vec n \cdot \vec y_s}a_{-\vec n}^*-
\e^{-\vec n \cdot \vec y_s}a_{\vec n}}{2\sinh(2\vec n \cdot \vec y_s)}
\quad\quad(\vec n \neq \vec 0)\,
\label{coeff1}   
\eea
as well as
\be
\lambda_{\vec 0}=\frac{1}{2}\,,\quad \mu_{\vec 0}=\frac{1}{2}\,
\label{coeff0}
\ee
and we can rewrite eq.(\ref{Ansatz}) in slightly simpler form as 
\be
P_{\vec n}(\vec y)\equiv\frac{\lambda_{\vec n}}{(2\pi\sigma)^{N/2}}
\exp\left(-\frac{(\vec y-\vec y_s)^2}{2\sigma}\right)+
\frac{\mu_{\vec n}}{(2\pi\sigma)^{N/2}}
\exp\left(-\frac{(\vec y+\vec y_s)^2}{2\sigma}\right)\,. 
\label{rewrite1}
\ee
Salcedo's construction \cite{Sa3,Sa4} is obtained by sending 
$\sigma\to 0$:
\be
P_{\vec n}(\vec y)= 
A_{\vec n}(\delta(\vec y-\vec y_s)+\delta(\vec y+\vec y_s))+
i B_{\vec n}(\delta(\vec y-\vec y_s)-\delta(\vec y+\vec y_s))\, 
\label{Ansatz'}
\ee
with 
\be
A_{\vec n}=\frac{\alpha_{\vec n}}{\cosh(\vec n \cdot \vec y_s)}\,,\qquad
B_{\vec n}=\frac{\beta_{\vec n}}{\sinh(\vec n \cdot \vec y_s)}\,.
\ee
$\vec y_s$ is still allowed to depend on $\vec n$ and may also be 
chosen different for the even and odd parts. Again this can be rewritten 
in slightly simpler form as
\be
P_{\vec n}(\vec y)\equiv \lambda_{\vec n} \delta (\vec y-\vec y_s)+
\mu_{\vec n}\delta (\vec y+\vec y_s)\,.
\label{rewrite'}
\ee 
with 
\bea
\lambda_{\vec n} =&\frac{\e^{\vec n\cdot \vec y_s}a_{\vec n}-
\e^{-\vec n \cdot \vec y_s}a_{-\vec n}^*} {2\sinh(2\vec n \cdot \vec y_s)}
\notag\\
\mu_{\vec n}=&\frac{\e^{\vec n \cdot \vec y_s}a_{-\vec n}^*-
\e^{-\vec n \cdot \vec y_s}a_{\vec n}}{2\sinh(2\vec n \cdot \vec y_s)}
\quad\quad(\vec n \neq \vec 0)\,.
\label{coeff2}
\eea
For the construction to be well-defined, it is obviously necessary that 
$\vec n \cdot \vec y_s \neq 0$ for all $\vec n$. One way to ensure this is 
by choosing $\vec y_s$ parallel to $\vec n$, i.e.
\be
\vec y_s =\alpha \vec n\,.
\label{parallel}
\ee
The damping of the higher modes is then manifest, but there is a problem 
with positivity: while by construction the coefficients $\lambda_{\pm\vec 
n}+\mu_{\mp\vec n}$ multiplying $\delta(\vec y\pm \alpha \vec n)$  are 
real, they are in general not positive. This can be cured, however, by 
spreading the zero mode over the different $\delta$ functions by setting: 
\be
P_{\vec 0}= \sum_{\vec n} r_{\vec n} \left(\delta(\vec y-\alpha \vec n)
+\delta(\vec y+\alpha \vec n)\right)
\ee
with
\be
\sum_{\vec n} r_{\vec n}=1/2\,.
\ee
Since the coefficients $\lambda_{\vec n},\,\mu_{\vec n}$ decay like
$\exp(-\alpha \vec n^2)$, it is possible to choose the $r_{\vec n}$ 
in such a way that the coefficients of $\delta(\vec y+\alpha \vec n)$
\be
r_{\vec n}+\lambda_{\vec n}+\mu_{-\vec n}>0\qquad 
\forall \vec n\neq \vec 0.
\ee  
We finally note that it is possible to rewrite the measure $P$, using the 
fact that multplication of the Fourier components corresponds to 
convolution in direct space, as follows:
\bea
P(\vec x,\vec y)=&\frac{1}{2(2\pi\sigma)^{N/2}}
\left[\exp\left(-\frac{(\vec y-\vec y_s)^2}{2\sigma}\right)
+ \exp\left(-\frac{(\vec y+\vec y_s)^2}{2\sigma}\right)\right]
\int d^N x'{\rm Re}\rho(\vec x')c(\vec x-\vec x')
\notag\\
+&\frac{1}{2(2\pi\sigma)^{N/2}}
\left[\exp\left(-\frac{(\vec y-\vec y_s)^2}{2\sigma}\right)
+ \exp\left(-\frac{(\vec y+\vec y_s)^2}{2\sigma}\right)\right]
\int d^N x'{\rm Im} \rho(\vec x')s(\vec x-\vec x')\,,
\label{convol1}
\eea
where convoluting smoothing functions $c$ and $s$ are
\be
c(\vec x)\equiv 1+
\sum_{\vec n \neq 0}\exp\left(-\frac{\vec n^2\sigma}{2}\right)
\frac{\cos(\vec n\cdot \vec x)}{\cosh(\vec n\cdot \vec y_s)};
\qquad
s(\vec x)\equiv 
\sum_{\vec n\neq 0}\exp\left(-\frac{\vec n^2\sigma}{2}\right)
\frac{\sin(\vec n\cdot \vec x)}{\sinh(\vec n \cdot \vec y_s)}\,.
\ee
For $\sigma = 0$ these formulae were given already in Ref.\cite{Sa4}, but 
because of the danger of the denominator $\sinh(\vec n \cdot \vec y_s)$ 
becoming small they were considered usable only for special cases. But as 
pointed out above, the problem can be avoided altogether by making the 
shift $\vec y_s$ dependent on $\vec n$, cf. eq.(\ref{parallel}) .

In the next section we will also encounter ``physical'' examples in which 
the formulation with a constant shift vector causes no problems, because 
the Fourier coefficients decay sufficiently fast to overcome the smallness 
of the denominator.
 
A further way of handling the difficulty of the small denominator 
$\sinh(\vec n \cdot \vec y_s)$ is worth pointing out:  we can make the 
shift $\vec y_s$ zero for the even part, and keep only the first order 
term in $\vec y_s$ in the odd part; in the latter case it is convenient 
to choose $\vec y_s$ parallel to $\vec n$. This way the even and odd 
parts become
\bea
P^+_{\vec n}(\vec y)=&\frac{A_{\vec n}}{(2\pi\sigma_+)^{N/2}}
\exp\left(-\frac{\vec y^2}{2\sigma_+}\right)
\notag\\
P^-_{\vec n}(\vec y)=&\frac{B_{\vec n}}{(2\pi\sigma_-)^{N/2}}
\vec y \cdot \vec n\exp\left(-\frac{\vec y^2}{2\sigma_-}\right)\,,
\label{noshift1}
\eea
where we also chose explicitly different width $\sigma_+$ and 
$\sigma_-$ for the even and odd parts. Consistency then requires 
\be
A_{\vec n}=\alpha_{\vec n} \exp(-\frac{\vec n^2\sigma_+}{2})\,,\quad 
B_{\vec n}=\frac{\beta_{\vec n}}{|\vec n| \sigma_-} \exp(-\frac{\vec
n^2\sigma_-}{2})\,.
\ee
where $\sigma_\pm$ may depend on $\vec n$. 

We can again express the resulting real density $P$ by convolution 
as follows: 
\bea
P(\vec x,\vec y)=&\frac{1}{(2\pi\sigma_+)^{N/2}}
\exp\left(-\frac{\vec y{\,^2}}{2\sigma_+}\right)
\int d^N x'{\rm Re}\rho(\vec x')c(\vec x-\vec x')
\notag\\+
&\frac{1}{(2\pi\sigma_-)^{N/2}} 
\exp\left(-\frac{\vec y^{\,2}}{2\sigma_-}\right) \left(\vec y \cdot 
\int d^N x'  {\rm Im} \vec\nabla_{\vec x} \rho(\vec x')\right)
s(\vec x-\vec x')\,.
\label{noshift2}
\eea
where the convoluting functions are now given by 
\be
c(\vec x)\equiv \sum_{\vec n} \cos(\vec n\cdot \vec x)
\exp(-\frac{\vec n^2\sigma_+}{2})= \frac{1}{(2\pi\sigma_+)^{N/2}}
\sum_{\vec n}\exp\left(-\frac{(\vec x-2\pi \vec n)^2}{2\sigma_+}\right)
\ee
and
\be
s(\vec x)\equiv
\sum_{\vec n\neq 0} \frac{1}{\sigma_-|\vec n|}
\cos(\vec n\cdot \vec x) \exp\left(-\frac{\vec n^2\sigma_-}{2}\right)\,.
\ee
To arrive at this we replaced $\vec n$ by $-i \vec \nabla_{\vec x}$ in the 
Fourier series and integrated by parts with respect to $\vec x$. 

The expression eq.(\ref{noshift2}) is manifestly real, but positivity is 
not 
obvious. In fact one has to choose $\sigma_-<\sigma_+$ and both quantities 
large enough. Another possibility is to replace the factor $\vec n 
\cdot \vec y$ in the odd part of eq.(\ref{noshift1}) by a bounded odd 
function of $\vec n \cdot \vec y$, for instance by setting
\be
P^-_{\vec n}(\vec y)=\frac{B_{\vec n}}{(2\pi\sigma_-)^{N/2}}
\tanh (\vec y \cdot \vec n)\exp\left(-\frac{\vec y^2}{2\sigma_-}\right)\,.
\ee 
The one can choose $\sigma_+=\sigma_-$.

The smoothing function $c$ is just the heat kernel on the torus
$\cT^N$. For real $\rho$ and $\vec y_s=\vec 0$ this 
becomes particularly simple:
\be
P(\vec x,\vec y)=\frac{1}{(2\pi\sigma_+)^{N/2}}
\left[\exp\left(-\frac{\vec y^2}{2\sigma_+}\right)
\int d^N x'{\rm Re}\rho(\vec x')c(\vec x-\vec x')\right]
\label{real}
\ee
The expression eq.(\ref{real}) can be recognized as a generalization of an 
example in Appendix B of \cite{aarts2017}.

%% file: sec3.tex
\section{ Examples}
\subsection{A prototype of a one plaquette action with the Polyakov line}  
\subsubsection{$\sigma > 0$}
The complex density reads 
\eqn
\rho_P(x)=\frac{1}{I_1(\beta)}e^{i x} \exp{\left( 
\beta\cos(x)\right)}, 
\label{rP}
\eqnx
with the Fourier components  
\eqn
a_n=\int_{-\pi}^{\pi} \frac{dx}{2\pi} e^{-i n x} \rho_P(x) 
= \frac{I_{n-1}(\beta)}{I_1(\beta)},  \;\;\;-\infty < n < \infty .\label{PF}
\eqnx
This leads to the manifestly real density.  Setting $y_s=1$ for 
simplicity, eq.(\ref{rewrite1}) gives
\bea
P_P(x,y)=&\frac{1}{2\sqrt{2\pi\sigma}}\left(e^{-(y-1)^2/2\sigma} + 
e^{-(y+1)^2/2\sigma} \right)\nn\\
+&\frac{1}{\sqrt{2\pi\sigma}}\sum_{n=1}^{\infty}e^{-n^2\sigma/2} 
\cos{(nx)} \left\{e^{-(y-1)^2/2\sigma} C_n^+ + e^{-(y+1)^2/2\sigma} 
C_n^-\right\},\nn\\
&C_n^+=\frac{e^n a_n-e^{-n}a_{-n}}{\sinh(2n)},\;\;\;\;\;\;\;\;\;\; 
C_n^-=\frac{e^{n} a_{-n}-e^{-n}a_{n}}{\sinh(2n)}\,,
\eea
which indeed becomes  positive for $\sigma \gtrsim 3.0$.

The weight $\rho_P(x)$ is normalized from the construction
\eqn
\int_{-\pi}^{\pi} \frac{dx}{2\pi} \rho_P(x)=a_0 = 1,
\eqnx
as well as the positive density
\eqn
\int_{-\pi}^{\pi}\frac{d x}{2\pi} \int_{-\infty}^{\infty} dy P_P(x,y) = 
\int_{-\pi}^{\pi}\frac{d x}{2\pi}\left\{1+\sum_{n=1}^{\infty} 
e^{-n^2\sigma/2}\left\{C_n^+ +  C_n^-\right\}\cos{(nx)}\right\}=1.
\eqnx

We then check two simple averages, $\la \cos(x) \ra$ and $\la 
\sin(x) \ra$. From the complex weight 
\eqn
\int_{-\pi}^{\pi} \frac{dx}{2\pi} \cos(x)\rho_P(x) 
=\frac{1}{2 I_1(\beta)}\left(I_0(\beta)+I_2(\beta)\right), \label{cavP}\\
\int_{-\pi}^{\pi} \frac{dx}{2\pi} \sin(x)\rho_P(x) 
=\frac{i}{2 I_1(\beta)}\left(I_0(\beta)-I_2(\beta)\right).\label{savP}\\
\notag
\eqnx
While for the positive density and with the aid of the following  simple 
gaussian integrals 
\eqn
\int_{-\infty}^{\infty} dy e^{i(x + i y)}e^{-(y \pm 
1)^2/2\sigma}=\sqrt{2\pi\sigma}e^{\pm 1+ i x+\sigma/2}\\
\int_{-\infty}^{\infty} dy e^{-i(x+i y)}e^{-(y \pm 
1)^2/2\sigma}=\sqrt{2\pi\sigma}e^{\mp 1-i x+\sigma/2}
\eqnx
one readily obtains
\eqn
\int_{-\pi}^{\pi}\frac{d x}{2\pi} \int_{-\infty}^{\infty} dy e^{i(x+i y)} 
P_P(x,y) = \frac{1}{2}\left(e^{-1}C_1^+ + e^1 C_1^-\right) 
=\frac{I_2(\beta)}{I_1(\beta)}\nn\\
\int_{-\pi}^{\pi}\frac{d x}{2\pi} \int_{-\infty}^{\infty} dy e^{-i(x+iy)} 
P_P(x,y) = \frac{1}{2}\left(e^{1}C_1^+ + e^{-1} C_1^-\right)=\frac{I_0(\beta)}{I_1(\beta)}
\eqnx
which indeed matches eq.(\ref{cavP}) and eq.(\ref{savP}).
Extension for arbitrary moments is straightforward.

\subsubsection{Polyakov line case at $\sigma=0$.} 

In this limit the positive density $P_P$ is just the sum of two Dirac 
delta distributions in $y$
\eqn
P_P(x,y)=P^+(x)\delta(y-y_s)+P^-(x)\delta(y+y_s)
\eqnx
with
\eqn
P^{\pm}(x)=\frac{1}{2} + \sum_{n=1}^{\infty} \cos{(n x)} C^{\pm}_n. \label{PPs}
\eqnx 
For given $\beta$ one can choose $y_s$  such that $P^{\pm}(x)$ are 
positive. For example, for $\beta \gtrsim 1.0$, $y_s \gtrsim 3.0$ is sufficient to 
guarantee dominance of the lowest mode for all $-\pi < x < \pi $, while 
for $\beta=0.1$ one needs $y_s \gtrsim 5.0$.

It may proove useful to rewrite  $P_P$ introducing an Ising-like variable 
$\sigma=\pm 1$ 
\eqn
P_P(x,y)=\sum_{\sigma=-1,+1}P^{(\sigma)}(x)\delta(y-\sigma y_s)
\eqnx
where
\eqn
C_n^{(\sigma)}=\frac{e^{n \sigma y_s} a_n-e^{-  n \sigma 
y_s}a_{-n}}{\sinh(2 n \sigma 
y_s)},\;\;\;\; 0 < n .
\eqnx
To illustrate the matching of both averages consider  $<\sin^2(x)>$.
Complex distribution gives
\eqn
\int_{-\pi}^{\pi} \frac{dx}{2\pi} \sin^2(x)\rho_P(x) 
=\frac{1}{4 I_1(\beta)}\left(I_1(\beta) - I_3(\beta)\right)
\eqnx
while using the positive density one obtains after a simple algebra
\eqn
\int \frac{dx}{2\pi} \sin^2{(x+i 
y_0)}\left\{\frac{1}{2}+\sum_{n=1}^{\infty} \cos{(n x)} C^+_n\right\}
+\int \frac{dx}{2\pi} \sin^2{(x-i 
y_0)}\left\{\frac{1}{2}+\sum_{n=1}^{\infty} \cos{(n x)} C^-_n\right\}=\nn\\
\frac{1}{2} - \frac{1}{4}\cosh{(2y_s)} \left(C^+_2+C^-_2\right)=
\frac{1}{4I_1(\beta)}\left(I_1(\beta)-I_3(\beta)\right), \nn
\eqnx
which is independent of $y_s$, and reproduces the above as expected.

\subsection{Complex one plaquette model with gauge invariance}

Again we consider the case $\sigma=0$.
The only nonvanishing, i.e. gauge invariant, complex density analogous to 
eq.(\ref{rP}) reads
\eqn
\rho_P(x_1,x_2,x_3,x_4)=
e^{i(x_1+x_2-x_3-x_4)}\exp{\left(\beta\cos(x_1+x_2-x_3-x_4)\right)} = 
\sum_{\vec{n}} a_{\vec{n}} e^{i \vec{n}\cdot\vec{x}} \nn 
\eqnx
the Fourier components are
\eqn
a_{\vec{n}}= \sum_m I_{m-1} 
\delta_{m,n_1}\delta_{m,n_2}\delta_{m,-n_3}\delta_{m,-n_4} \label{f4}
\eqnx 
The positive distributions eq.(\ref{PPs}) can be written in a manifestly 
real 
form
\eqn
P^+(\vec{x})=
\frac{ a_{\vec{0}}}{2}+\sum_{\vec{n},\vec{n}\ne\vec{0}} \frac{e^{ 
\vec{n}\cdot \vec{y}_s } a_{\vec{n}}}{\sinh{(2\vec{n}\cdot \vec{y}_s 
)}}\cos{\left({ \vec{n} \cdot \vec{x}}\right)},
\eqnx
and similarly for $P^-$ contribution.

With the explicit form of the Fourier components eq.(\ref{f4}) we choose 
the symmetric shift vector, $\vec{y}_s=y_s(1,1,-1,-1)$, to obtain for both 
components $(\sigma=\pm1) $
\eqn
P^{\sigma}(\vec{x})=\frac{ I_1}{2}+\sum_{m, m \ne 0} 
 \frac{e^{ 4 m y_s } I_{m-\sigma}}{\sinh{( 8 m y_s )}}\cos{\left( m 
(x_1+x_2-x_3-x_4\right)}.
\eqnx 
Not surprisingly, the final result is very much the same as for the one 
variable with the replacement
$ x \rightarrow x_P=x_1+x_2-x_3-x_4$ and a multiple 
$\delta^{(4)}(\vec{y}-\sigma\vec{y}_s)$ function. 

\subsection{2x2 periodic lattice, with two Polyakov lines}

\begin{figure}[h]
\begin{center}
\includegraphics[width=6cm]{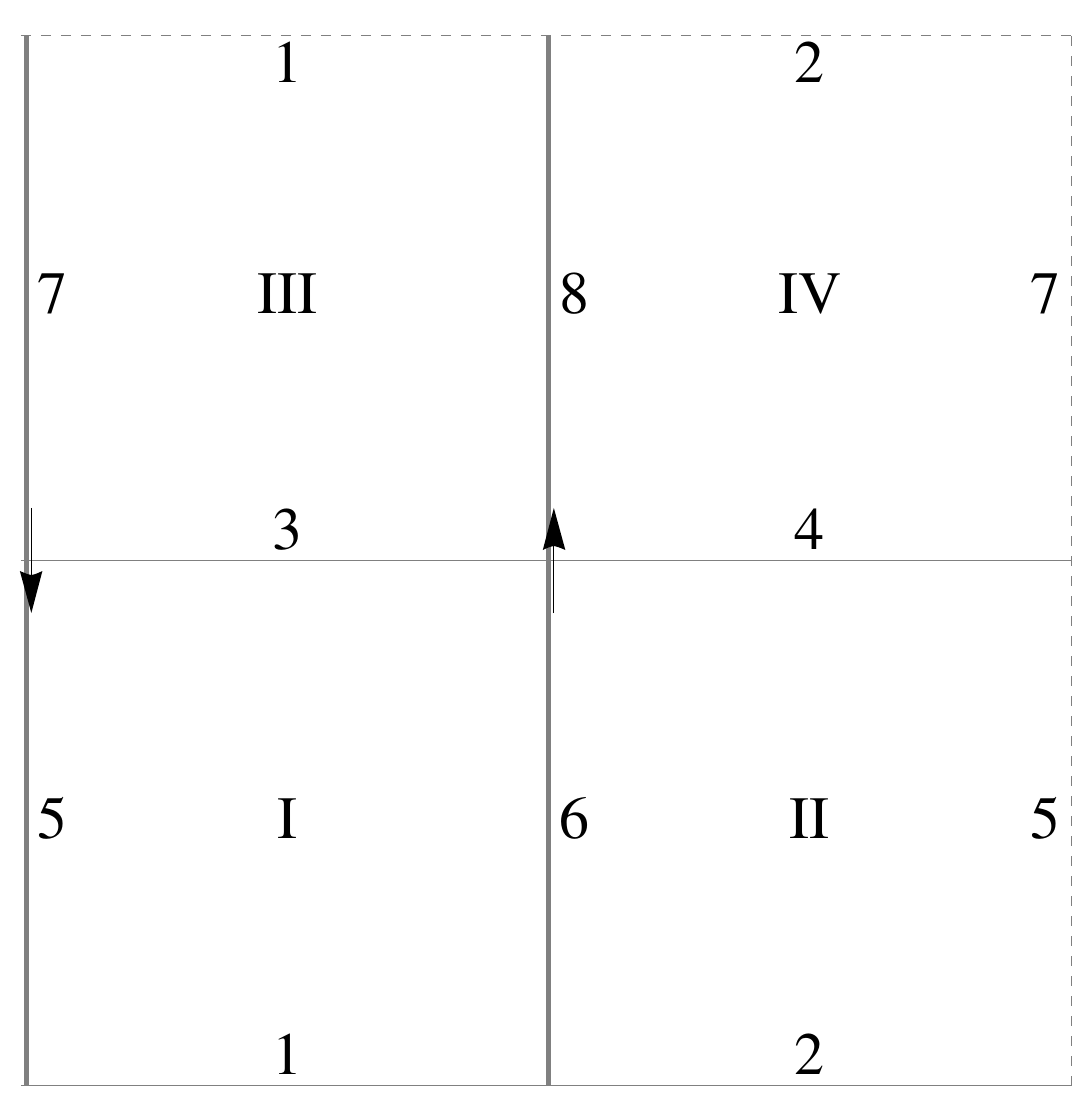}
\end{center}
\vskip-4mm \caption{A $2 \times 2$ lattice with two Polyakov lines.} 
\label{fig:f1}
\end{figure}

The complex density depends now on 8 link angles labeled as in 
figure\ref{fig:f1}.
To save writing we sometimes denote a component by its mere number i.e. 
$x_5\rightarrow 5$
\bea
\rho(\vec x)=&B(3+8-1-7)B(4+7-2-8)\nn\\
                    &B(1+6-3-5)B(2+5-4-6)\nn\\
                    &U(-5-7)U(6+8)\nn\\
\eea
where
\be
B(\phi)=e^{\beta\cos{(\phi)}}\,, U(\phi)=e^{i \phi}\label{2x2C}
\ee
Many link variables are redundant due to the gauge invariance. However in 
two dimensions one can rewrite the partition function, up to a constant 
Jacobian, in terms of independent plaquette angles $\vec{\phi}$
\eqn
Z_{\phi}=\int_{\vec{\phi}} \rho(\vec{\phi}).
\eqnx
For a 2x2 lattice there are three independent plaquettes and we choose 
$(\phi_{I},\phi_{II},\phi_{III})\equiv(\phi_{1},\phi_{2},\phi_{3})$  as 
our variables
\eqn
\rho(\vec{\phi})= B(\phi_1) B(\phi_2)B(\phi_3) B(\phi_1+\phi_2+\phi_3) 
U(\phi_1) U(\phi_3). \label{BP}
\eqnx
The only reason this system does not factorize is due to the constraint 
$\phi_1+\phi_2+\phi_3+\phi_4=0$, forced by the periodic boundary 
conditions.

The Fourier components are then 
\bea
a_{\vec{n}}=&\int_{-\pi}^{\pi} \frac{d\phi_1}{2\pi} e^{-i 
n_1\phi_1}\int_{-\pi}^{\pi} \frac{d\phi_2}{2\pi} e^{-i 
n_2\phi_2}\int_{-\pi}^{\pi} \frac{d\phi_3}{2\pi}e^{-i n_3\phi_3} \nn\\
&e^{\beta\cos(\phi_1)}e^{\beta\cos(\phi_2)} 
e^{\beta\cos(\phi_3)}e^{\beta\cos(\phi_1+\phi_2+\phi_3)} 
e^{i(\phi_1+\phi_3)}=\nn\\
&\sum_m I_m I_{m-n_2} I_{m-n_1+1} I_{m-n_3+1}
\eea
and the corresponding positive density can be readily constructed. 
Replacing $\phi_k$ by $x_k+i y_k$, we get
\bea
P(\vec{x},\vec{y})=&
\frac{a_{\vec{0}}}{2}\left( 
\delta(\vec{y}-\vec{y}_s)+\delta(\vec{y}+\vec{y}_s) \right)\nn\\
+&\sum_{\vec{n}\ne\vec{0} } 
 \left\{\frac{e^{\vec{n}\cdot\vec{y}_s}a_{\vec{n}}-
e^{-\vec{n}\cdot\vec{y}_s}a_{-\vec{n}}}
{2\sinh{(2 \vec{n}\cdot\vec{y}_s)}}\delta(\vec{y}-
\vec{y}_s)+\frac{e^{\vec{n}\cdot\vec{y}_s}a_{-\vec{n}}-
e^{-\vec{n}\cdot\vec{y}_s}a_{\vec{n}}}
{2\sinh{(2 \vec{n}\cdot\vec{y}_s)}}\delta(\vec{y}+\vec{y}_s)\right\} 
e^{i \vec{n}\cdot\vec{x}}
\eea
To guarantee that zero of sinh is only at the origin, we take 
$\vec{y}_s=y_s (1,\sqrt{2},\sqrt{3})$. Again choosing large enough $y_s$ 
makes the lowest mode dominant, leading to positivity of $P$.  

{\it Remark:} There is a mathematical (number theoretic) subtlety behind 
this: while it is obvious that the denominator can never vanish, it will 
nevertheless become very small whenever the integer vector $\vec n$ becomes 
very nearly parallel to a real vector $\vec y$ with $\vec y\cdot \vec 
y_s=0$. Numerically we see that this happens only for $|\vec n|$ so large 
that the decay of the Fourier coefficients $a_{\vec n}$ compensates for the 
small denominator.

\begin{figure}[h]
\includegraphics[width=8cm]{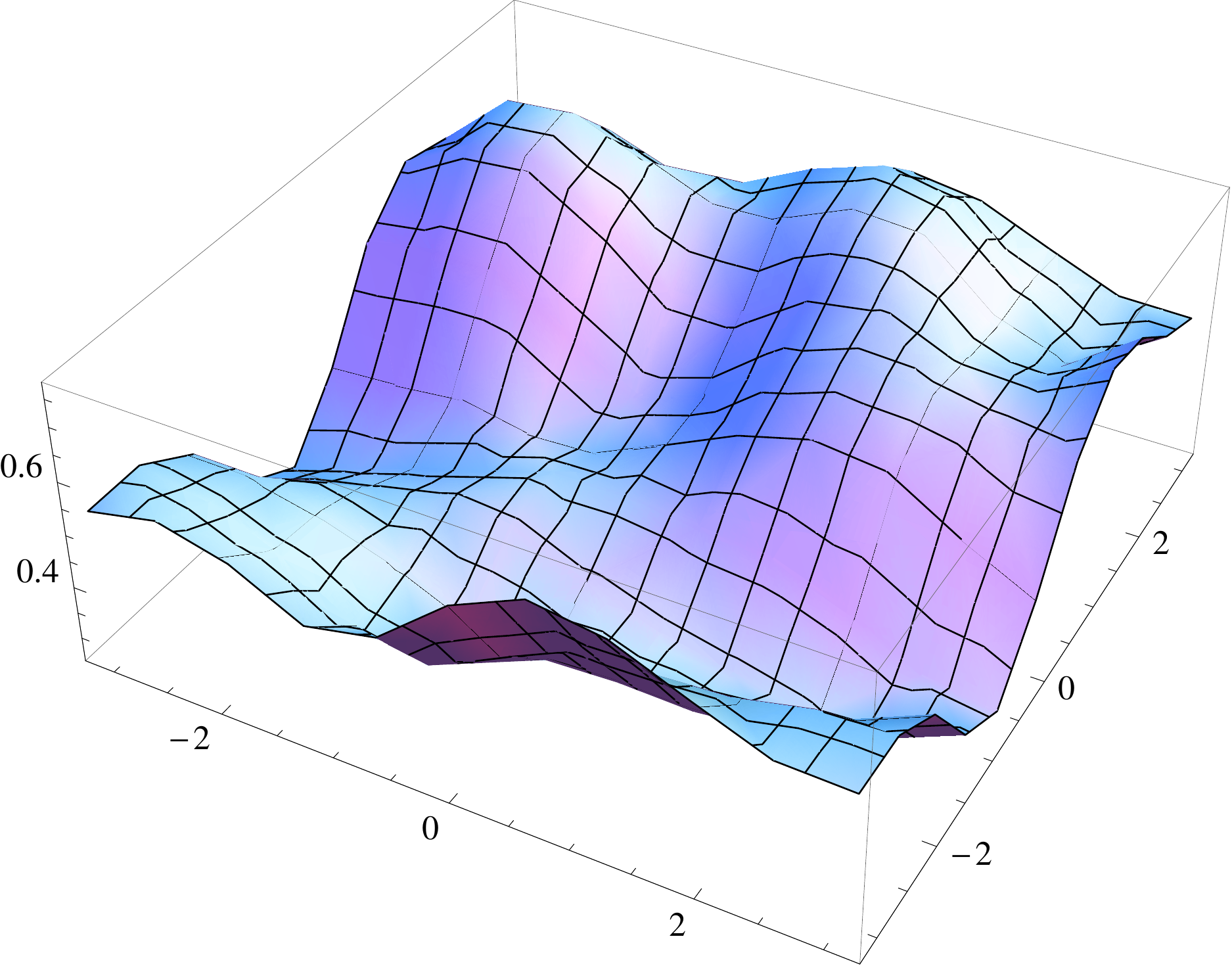}\hspace*{1cm}
\includegraphics[width=8cm]{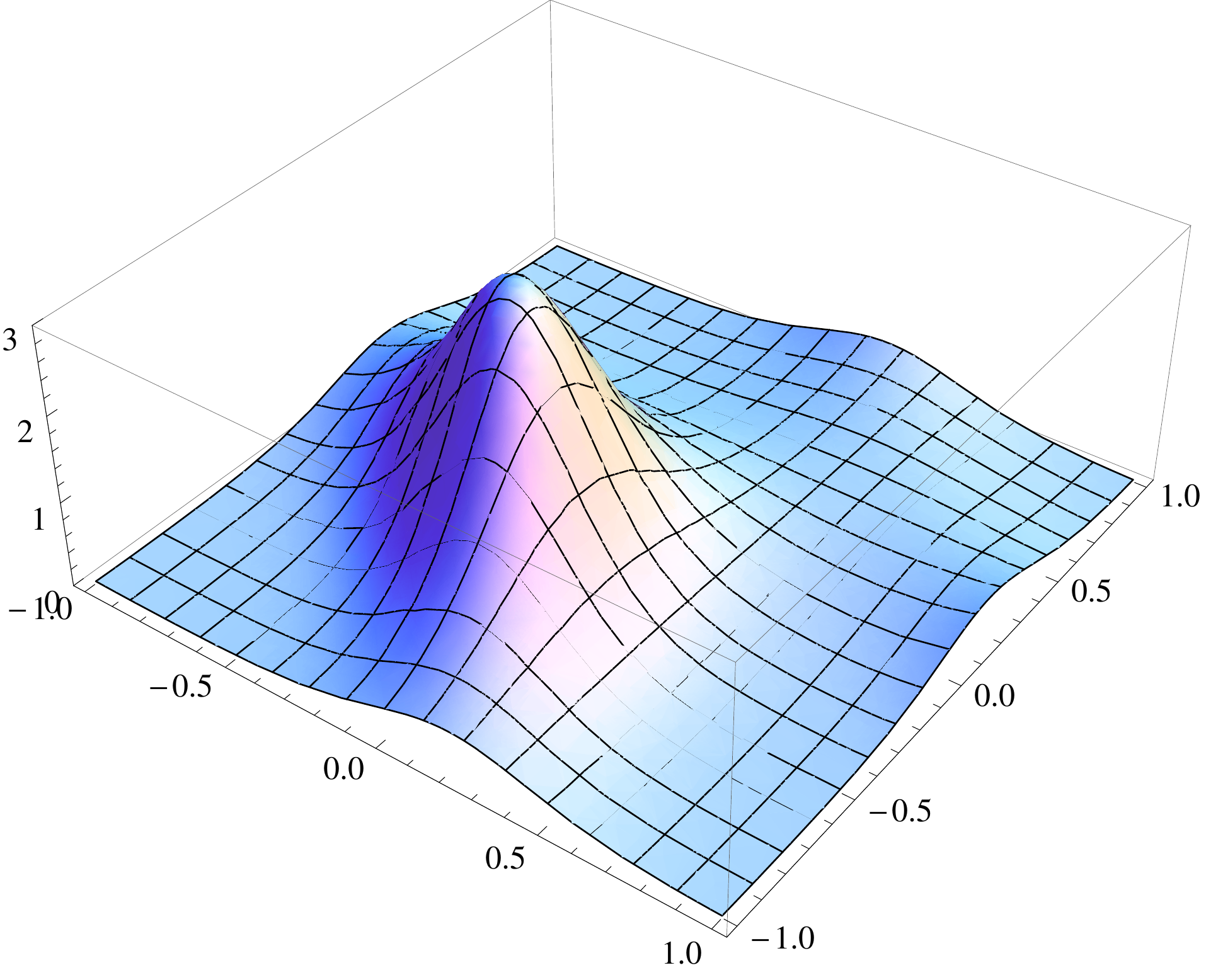}
\vskip-4mm \caption{Positive densities with (left) and without the 
Polyakov loop (right).} \label{fig:f2}
\end{figure}

A particular cross section $-\pi < x_1 < \pi,\;-\pi < x_2 < \pi,\; $of 
$P^+(\vec{x})$ at $x_3=.6\pi$ and $y_s=8.0$ is displayed in figure 
\ref{fig:f2}. To see the effect of Polyakov lines we show also the 
original positive Boltzmann weight eq.(\ref{BP}) without the loop factors. 
Needless to say, the effect of Polyakov lines is dramatic, emphasizing the 
relevance of the statistical formulation.

The second good news is that the variation of $P$ is substantial. This 
means that the dominance of the first mode, required for positivity, does 
not preclude the importance of other modes giving rise to a nontrivial 
structure of the positive distribution. 


Instead of checking a particular simple observable, we can verify that the 
matching equations are satisfied for arbitrary moments of three independent 
U(1) variables.

It is readily seen that the normalisation of both densities indeed coincide
\eqn
\int \frac{d^3\phi}{(2\pi)^3}\rho(\vec{\phi})=\int \frac{d^3 
x}{(2\pi)^3}d^3 y P(\vec{x},\vec{y})\equiv Z=a_{\vec{0}}
\eqnx
It remains to verify the averages of generic moments
\eqn
\la \left(e^{i \phi_1}\right)^{r_1}  \left(e^{i \phi_2}\right)^{r_2}  
\left(e^{i \phi_3}\right)^{r_3} \ra_{\rho(\vec{\phi})} = \la \left(e^{i 
(x_1+i y_1)}\right)^{r_1} \left(e^{i (x_2+i y_2)}\right)^{r_2}  \left(e^{i 
(x_3+i y_3)}\right)^{r_3} \ra_{P(\vec{x},\vec{y})} 
\eqnx
The LHS is obviously the Fourier component $a_{-\vec{r}}$. 

On the other hand, the positive distribution $P$ reproduces the Fourier 
modes by its very construction. Let us check this nevertheless.
\bea
&Z\la \left(e^{i (x_1+i y_1)}\right)^{r_1} \left(e^{i (x_2+i 
y_2)}\right)^{r_2}  \left(e^{i (x_3+i y_3)}\right)^{r_3} 
\ra_{P(\vec{x},\vec{y})} =\nn\\
&\int d^3 y e^{-\vec{r}\cdot\vec{y}}\int 
\frac{d^3 x}{(2\pi)^3} e^{i \vec{r}\cdot\vec{x}}
\Biggl(\frac{a_{\vec{0}}}{2}\delta(\vec{y}-\vec{y}_s)+
\frac{a_{\vec{0}}}{2}\delta(\vec{y}+\vec{y}_s)
+\nn\\ &\sum_{\vec{n}\ne\vec{0} } 
e^{i \vec{n}\cdot\vec{x}}\left\{\frac{e^{\vec{n}\cdot\vec{y}_s}a_{\vec{n}}-
e^{-\vec{n}\cdot\vec{y}_s}a_{-\vec{n}}}{2\sinh{(2 \vec{n}\cdot\vec{y}_s)}}
\delta(\vec{y}-\vec{y}_s)+
\frac{e^{\vec{n}\cdot\vec{y}_s}a_{-\vec{n}}-e^{-\vec{n}\cdot\vec{y}_s}
a_{\vec{n}}}{2\sinh{(2 \vec{n}\cdot\vec{y}_s)}}
\delta(\vec{y}+\vec{y}_s)\right\}\Biggr)\nn\\
&a_{\vec{0}}\delta_{\vec{r}}+  
(1-\delta_{\vec{r}})\left(e^{-\vec{r}\cdot\vec{y}_s} 
\frac{e^{-\vec{r}\cdot\vec{y}_s}a_{-\vec{r}}- 
e^{\vec{r}\cdot\vec{y}_s}a_{\vec{r}}}{2\sinh{(-2 \vec{r}\cdot\vec{y}_s)}}+
e^{\vec{r}\cdot\vec{y}_s}\frac{e^{-\vec{r}\cdot\vec{y}_s}a_{\vec{r}}- 
e^{\vec{r}\cdot\vec{y}_s}a_{-\vec{r}}} {2\sinh{(-2 
\vec{r}\cdot\vec{y}_s)}}\right) =a_{-\vec{r}}\nn
\eea
  
\subsection{Larger lattices and separability in plaquette variables}

\begin{figure}[h]
\begin{center}
\includegraphics[width=6.3cm]{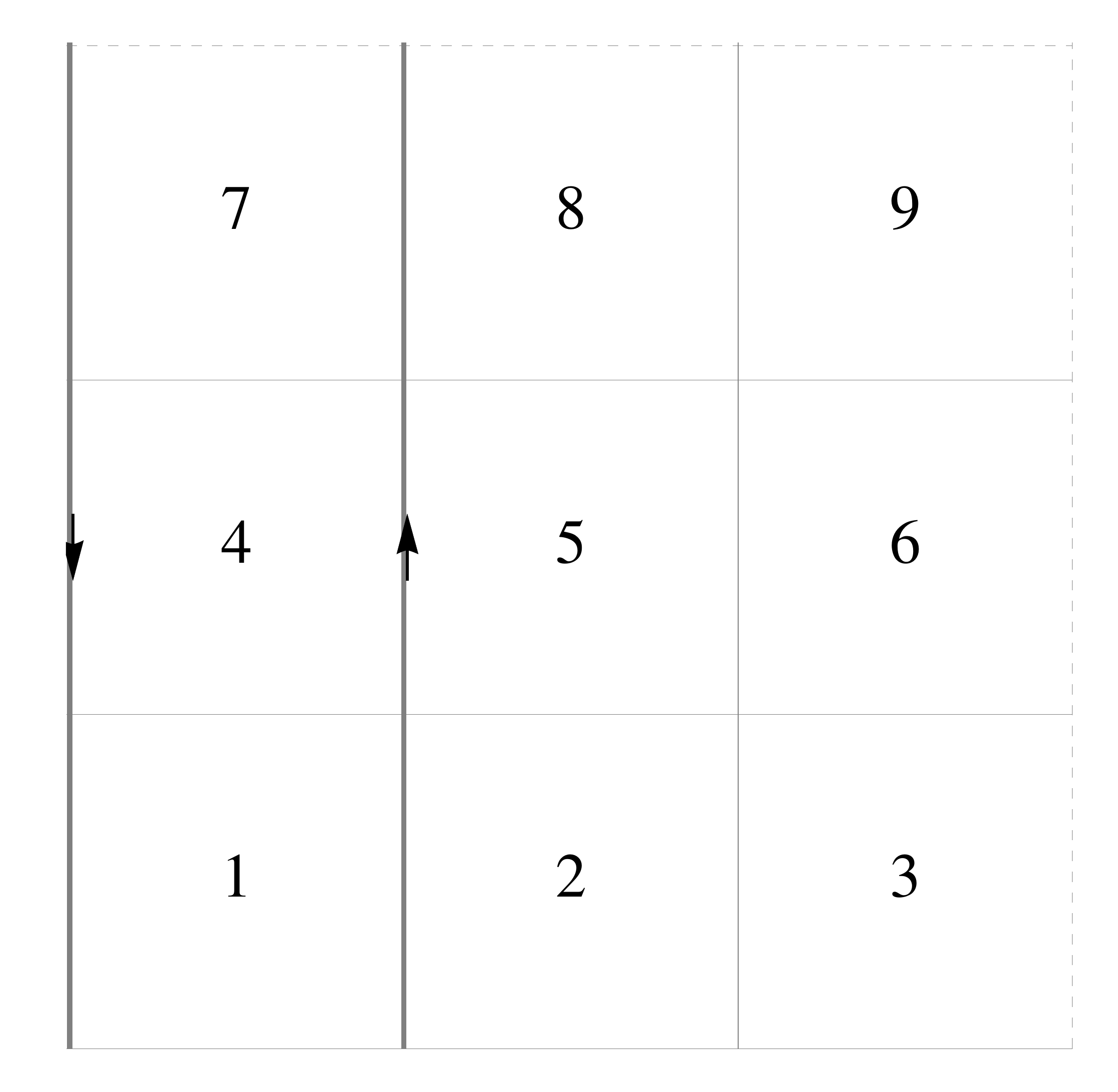}\hspace*{1cm}
\includegraphics[width=6cm]{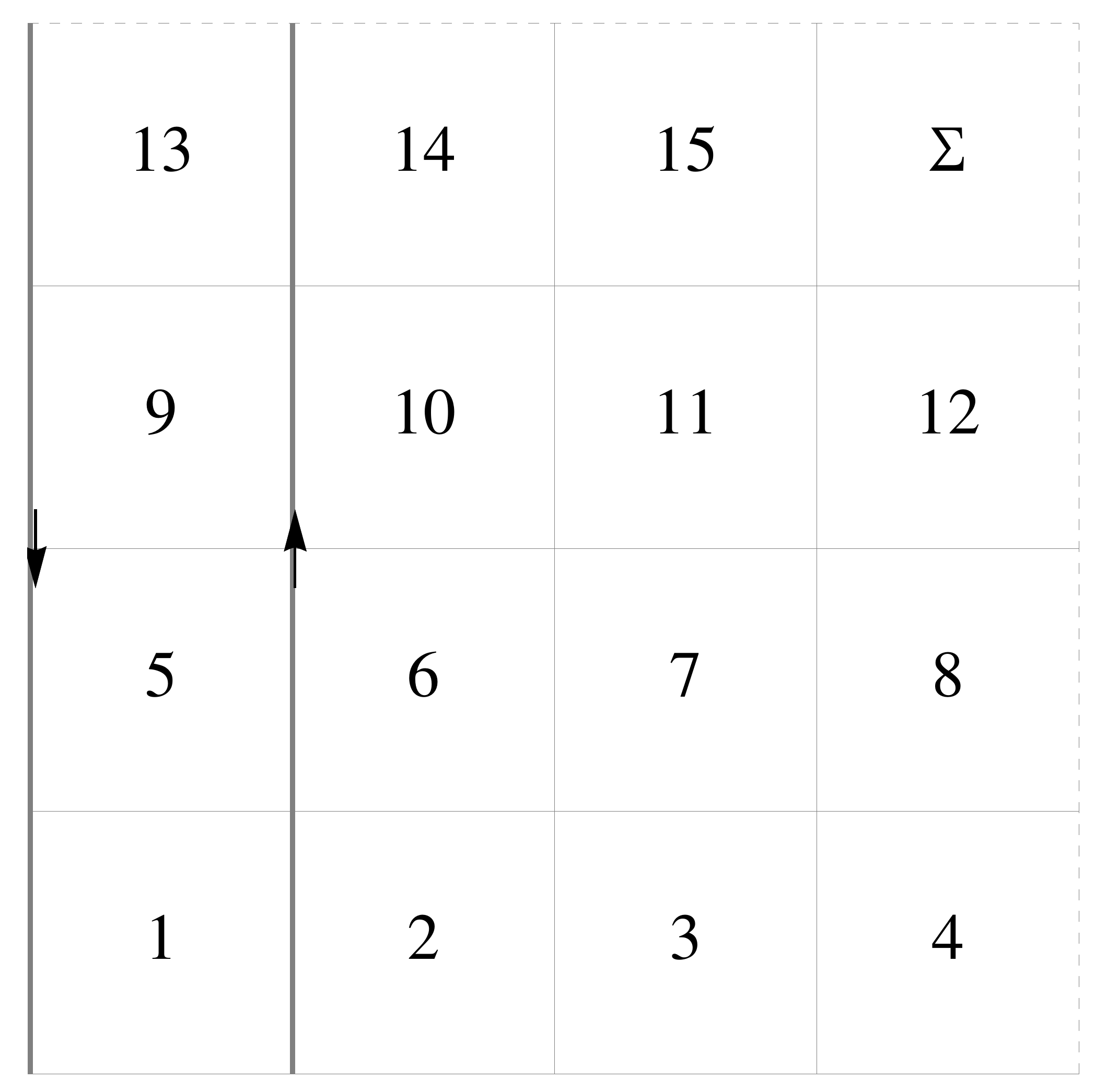}
\end{center}
\vskip-4mm \caption{Towards larger abelian lattices. Plaquette labeled Σ 
$\Sigma$ is determined by the others and periodic boundary conditions.} 
\label{fig:f3}
\end{figure}

Generalization to larger lattices is straightforward, c.f. figure 
\ref{fig:f3}.

For example the complex weight which includes two Polyakov lines on the 
3x3 periodic lattice reads
\eqn
\rho(1...8)=B(1)B(2)...B(8)B(1+2+3+...+8)U(1)U(4)U(7) \label{rho3}
\eqnx
and its Fourier components are readily available
\be
a_{\vec{n}}= \sum_m  
(I_{m-n_7+1} I_{m-n_8}I_{m})\,
(I_{m-n_4+1} I_{m-n_5}I_{m-n_6})\,
(I_{m-n_1+1} I_{m-n_2}I_{m-n_3}) \label{F33}
\ee
and similarly for larger, $N \times N$, lattices. The rule for arbitrary 
lattice is evident: the Fourier coefficients are the sums, over one 
flux $m$, of products of $N^2$ modified Bessel functions. The indices of 
Bessel functions are given by the differences $m-n_i$ if corresponding 
plaquettes do not tail inner region between the Polyakov lines and they are 
shifted by one if they do. The index of the last Bessel function is just 
$m$. All this is well known e.g. from the duality transformations studied 
decades ago \cite{elitzur,banks}.

Corresponding positive distributions follow from our general 
representation eq.(\ref{rewrite1}). 

Before concluding this section let us discuss briefly the issues of 
locality and separability of resulting positive distributions in the case 
of many variables.

In practical applications it is desirable, but not necessary, that 
$P(\vec{x},\vec{y})$ admits a local update algorithm. For example the 
original density eq.(\ref{rho3}) without Polyakov lines does so. Even the 
"last" Boltzmann factor can be easily accommodated by a local algorithm. 
Alternatively one can employ a free boundary conditions or just ignore the 
single constraint in the large $N$ limit. In this case our complex density 
reads
\eqn
\rho(\vec{\phi})=\left(\prod_{exterior\; of\; W} B(\phi_{ext})\right) 
\left( \prod_{interior\; of\; W} B(\phi_{int}) U(\phi_{int}) \right)
\eqnx
This does not only factorize, but separates as well, i.e. no common 
variables are shared among these factors. Therefore our construction can 
be applied independently to each factor.

However in higher dimensions there are many more relations between 
plaquette variables, even for free boundary conditions, so the positive 
measures will be highly nonlocal and one has to look for other 
possibilities.

Second, with more variables the direct Fourier inversion  becomes much 
more time consuming. Hence one should explore other representations which 
would allow resummation of the series eq.(\ref{fourier}).

It remains to be seen to what extent the freedom inherent in the whole 
approach can be used to circumvent these problems and to generalize the 
method to more interesting, higher dimensional systems.

%% file: sec4.tex
\section{Summary and directions of future research}

Given a complex measure on an $N-$torus, we have presented a general 
construction of a positive (probability) measure on the complexification of 
that torus, which is equivalent in the sense that expectations of 
holomorphic observables agree. We identified necessary and sufficient 
conditions which such a measure has to fulfill (in addition to being 
positive) and tested the construction in a number of cases.

Our examples range from one to many degrees of freedom, also with local gauge invariance.
They all focus on the classic, by now, question of the space structure of confining strings. It was attracting considerable attention from the beginning of lattice field theory. However the problem was always plagued by strong oscillations of the measure induced by Wilson lines. In the last application the positive density, which incorporates Wilson lines, is explicitly constructed. Even though this is achieved for the simple abelian, and lower dimensional, system the intricacies of the large number of variables were successfully addressed, and exposed in this context, for the first time.

We did not present in this paper an algorithm capable of dealing with 
complex actions on large lattices. The reason is that the positive 
measures we are able to construct, are not local in the sense that they 
are not built up from contributions located at links, plaquettes or 
other small structures. Our aim was so far to demonstrate mathematically 
that positive measures representing complex ones exist quite 
generally, and to provide an explicit construction for them.

With respect to their inherent nonlocality, our measures should be 
compared to the effective measures for theories with fermions featuring a 
nonlocal determinant of the Dirac operator (and zero chemical potential). 
In that case, eventually a workable algorithm was found: the 
universally used `Hybrid Monte Carlo' algorithm. So one direction of 
research to be pursued is to search for the possibility to overcome the 
nonlocality obstacle. Let us list several avenues that could be followed:

1) Search for an analogue of the `Hybrid Monte Carlo' algorithm for our
    measures.

2) Since the key of the construction is to decompose the complex measure
    using harmonic (Fourier) analysis, Fast Fourier Transform (FFT) could
    be helpful.

3) Applying the Real Langevin algorithm to our positive measures appears
    to be feasible, just as the Complex Langevin algorithm can be applied
    to full QCD with chemical potential \cite{sexty,aarts2017}. For toy
    models the Real Langevin algorithm has been applied successfully, but
    this is rather trivial and we do not want to burden the reader with details
    about such a well-known method.

4) Salcedo \cite{Sa3} has devised an ingenious scheme to deal with the
    nonlocality issue by regarding each variable to be updated separately.
    The original complex measure, keeping all other variables
    fixed, is considered as a measure for only the variable to be updated;
    by complexifying only the variable in question, one obtains an
    equivalent positive measure for this variable, which can then be
    updated by a standard Monte Carlo procedure. The feasibility of this
    strategy for larger lattices is yet to be tested.

Furthermore there is the issue of generalizing the construction to 
non-abelian gauge theories. A straightforward generalization would use 
harmonic analysis for non-abelian groups. Some results in this direction 
have been recently obtained in Ref. \cite{Sa5}. 
\vskip1cm
Acknowledgment: this work is supported in part by the NCN grant 
UMO-2016/21/B/ST2/01492.